# On Maximizing Coverage in Gaussian Relay Networks

Vaneet Aggarwal
Department of Electrical Engineering
Princeton University
Email: vaggarwa@princeton.edu

Amir Bennatan
Program for Applied and
Computational Mathematics
Princeton University
Email: abn@math.princeton.edu

A. Robert Calderbank
Department of Electrical Engineering
Princeton University
Email: calderbk@math.princeton.edu

*Abstract*—Results for Gaussian relay channels typically focus on maximizing transmission rates for given locations of the source, relay and destination. We introduce an alternative perspective, where the objective is maximizing *coverage* for a given rate. The new objective captures the problem of how to deploy relays to provide a given level of service to a particular geographic area, where the relay locations become a design parameter that can be optimized. We evaluate the decode and forward (DF) and compress and forward (CF) strategies for the relay channel with respect to the new objective of maximizing coverage. When the objective is maximizing rate, different locations of the destination favor different strategies. When the objective is coverage for a given rate, and the relay is able to decode, DF is uniformly superior in that it provides coverage at any point served by CF. When the channel model is modified to include random fading, we show that the monotone ordering of coverage regions is not always maintained. While the coverage provided by DF is sensitive to changes in the location of the relay and the path loss exponent, CF exhibits a more graceful degradation with respect to such changes. The techniques used to approximate coverage regions are new and may be of independent interest.

## I. INTRODUCTION

Relay channels have recently attracted significant attention as a model for *ad-hoc* networks [7]. These channels model problems where one or more *relays* help a pair of terminals communicate. The general channel model was first considered by van der Meulen [1], [2], [3] and further studied in a ground breaking work by Cover and El Gamal [4]. Although the capacity region for the channel is still unknown, the results of [4] include two achievable coding strategies which were subsequently named decode-and-forward (DF) and compress-and-forward (CF). The Gaussian relay channel was examined by Kramer *et al.* [5] and Høst-Madsen and Zhang [8].

A distinctive property of the existing results in the literature is that they all consider the classic information-theoretic perspective of maximizing the achievable *rate* for given locations of the source, relay and destination nodes. However, in many cases of practical interest, the design problem at hand is to maximize *coverage* for a fixed desired transmission rate. This is the focus of our paper.

The following simple example, illustrates the difference between the performance measure considered in this paper (maximizing coverage), and the classic measure (maximizing rate). Consider a source, relay and destination all at equal distances (say 1) from one another (on the vertices of an equilateral triangle), as in Fig. 1. Assume equal power constraints ($P_1 = P_2 = 1$). Assume the requisite rate is $R = 1$ bits/channel use. Then DF can achieve a maximum rate of 1 bit/channel use at the destination, yet CF can do better: 1.17. In this paper (Theorem 1, Sec. III) we assert that whenever the relay can decode, DF provides superior performance to that of CF. This example would appear to contradict that assertion. However, with the performance measure of our paper (coverage), the advantage of CF does not matter. This is because we are not concerned with the maximum achievable rate, as long as it is greater than $R$.

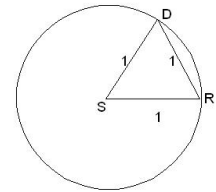

Fig. 1. Locations of the source, the relay and the destination in the above example.

In their seminal work, Kramer *et al.* [5][Sec. VII.B] extended the classic treatment of the relay problem by considering the location of the relay as a design parameter. That is, they considered the effect of relocating the relay on the achievable rates at the destination.

This effect is heavily dependent on the *destination's* location. While a destination at some locations may benefit from relocating the relay, a destination at other locations will suffer. However, in many cases of practical interest, the location of the relay is determined at a time when the destination's location is unknown. The destination is typically a mobile station, while the relay is often a fixed terminal, whose location is determined once, at the time that the network is designed.

Nevertheless, while the location of the destination may be unknown at the time of network design, the target transmission rate is typically known. Thus, the effect of changes in the relay (change of location or communication strategy) on the *coverage region* can be evaluated.

Our analysis further extends the discussion of [5] in the following ways,

1) The discussion of [5] is limited to a single destination, at a fixed distance of 1 (normalized distance metric) from the source, and to a relay that is located on the line segment connecting the source to the destination. Our discussion is completely general.
2) In [5][Remark 31], the authors analyze the performance of DF and CF in the limit when the relay is either close to the source or close to the destination. They provide numerical data when the relay is intermediate these two extremes. In this paper, we provide rigorous analysis for all possible locations of the relay.
3) The results of [5] provide the following intuition: Whenever the channel from the source to the relay is strong enough to enable the relay to decode the source's message, DF renders superior performance to that of CF. However, regardless of how strong (or weak) this channel may be, there is always *some* rate (however low) that the channel can support (under the channel model both paper share). Strictly speaking, with the formulation of [5], the relay can *always* decode, and thus the intuition cannot be stated formally. The introduction of a target transmission rate, in this paper, enables us to formalize and prove the intuition.

We begin in Sec. II by providing some background on the channel model and achievable strategies for it. We also formally define the concept of coverage. In Sec. III we compare the coverage regions of the CF and DF achievable strategies, for different locations of the relay. In this comparison, we assume that the relay's location is the same with both strategies. We extend the comparison in Sec IV, and allow each strategy its own preferred relay location. In Sec. V we provide bounds on the area (measured in normalized units of area) of the coverage region of DF, as a function of the relay's location. The discussion in this paper mostly focuses on a full-duplex non-fading channel model. In Sec. VI we briefly discuss additional channel models. Sec. VII concludes the paper. Throughout the paper, proofs are deferred to the appendix.

## II. BACKGROUND AND DEFINITIONS

### A. The Channel Model

In this section, we introduce a simple relay channel model, which will be our focus throughout most of the paper. In this model, nodes are assumed to be full-duplex. This means that a node can receive and transmit simultaneously. Furthermore, the signal attenuation between any two points assumed to be a deterministic function of the distance between the two. In Sec. VI we will consider additional channel models (random fading and half-duplex).

Our model is depicted in Fig. 2. The channel consist of three nodes: a source (node 1), a relay (node 2) and a destination (node 3). We consider a two-dimensional domain for our three-node network. This means that the source, relay and destination are associated with two-dimensional location vectors $\mathbf{a}_1$, $\mathbf{a}_2$ and $\mathbf{a}_3$, respectively. For simplicity, and without loss of generality, we may assume that $\mathbf{a}_1 = (0,0)$, and $\mathbf{a}_2 = (d,0)$ where $d > 0$ is the distance between the source and the relay. The relations between the channel outputs and inputs are a function of the distances between the various nodes. We let $d_{kl}, k,l = 1,2,3$ denote the distances between nodes $k$ and $l$. With this notation, $d_{12} = \|\mathbf{a}_2\| = d, d_{13} = \|\mathbf{a}_3\|$ and $d_{23} = \|\mathbf{a}_3 - \mathbf{a}_2\|$ where $\|\cdot\|$ denotes the Euclidean norm. See Fig. 2. The channel equations are now given by,

$$y_2[i] = \frac{1}{d_{12}^{\alpha/2}}x_1[i] + z_1[i]$$

$$y_3[i] = \frac{1}{d_{13}^{\alpha/2}}x_1[i] + \frac{1}{d_{23}^{\alpha/2}}x_2[i] + z_2[i]$$

where, $x_1[i]$ and $x_2[i]$ are the signals transmitted from the source and relay, respectively, at time $i$. These signals are subject to average power constraints $P_1$ and $P_2$, respectively. $y_2[i]$ and $y_3[i]$ denote the observed signals at the relay and destination, respectively. $z_1[i]$ and $z_2[i]$ are mutually independent i.i.d circularly-symmertic complex Gaussian noise with variance 1. $\alpha \geq 2$ is the path loss exponent.

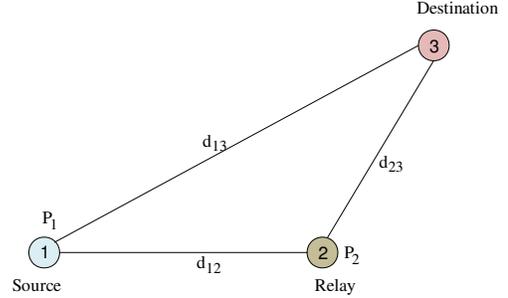

Fig. 2. Single Relay Network

### B. Codes and Achievable Strategies

A code for the relay channel of rate $R$ and block-length $n$, consists of a pair $(C, \{f_i\}_{i=1}^n)$. $C$ is a set of $2^{nR}$ codewords of length $n$. The source encoder constructs its signal by selecting a codeword $\mathbf{x}_1 \in C$. At time $i$, the source sends index $x_1[i]$ and the relay sends $x_2[i]$ using $f_i$ based on its past observation. That is, $x_2[i] = f_i(y_2[i-1], ..., y_2[1])$.

A relay transmission *scheme* $\mathcal{S}$ is formally a collection of relay codes.

*Definition 1:* Given locations $\mathbf{a}_2 = (d,0)$ and $\mathbf{a}_3$ of the relay and the destination respectively, a rate $R$ is defined to be *achievable* by a scheme $\mathcal{S}$ (equivalently: $\mathcal{S}$ *supports* $R$) if for any $\epsilon > 0$, there exist $(C, \{f_i\}_{i=1}^n) \in \mathcal{S}$ such that the rate of $C$ is at least $R$, and the probability of error, under maximum-likelihood decoding is at most $\epsilon$.

We define the capacity of $\mathcal{S}$ at relay location $\mathbf{a}_2 = (d,0)$ and destination location $\mathbf{a}_3$ as,

$$C_\mathcal{S}(d, \mathbf{a}_3) = \sup\{R : \mathcal{S} \text{ supports } R\}$$

Cover and El Gamal [4] introduced two achievable coding strategies which were subsequently named decode-and-forward (DF) and compress-and-forward (CF). With DF, the relay decodes the message transmitted by the source. It then cooperates with the source to transmit the message to the destination. With CF, the relay considers the observed signal from the source as a raw signal, compresses it, and transmits it to the destination. The destination then combines this observation with its own observation, and uses both to decode the source's message.

An important distinction between the two strategies, is that with DF, the relay attempts to decode the source's message, while with CF it does not. A comprehensive description of the strategies is available e.g. in [4], [5], [8].

The achievable rates with both schemes, for our channel model, were computed in [5], [8], and are provided in Appendix A. Following their example, we confine our attention to CF when the random variables used in the generation of the codebooks are Gaussian.

*C. Coverage*

We are now ready to formally define the concept of coverage.

*Definition 2:* Let $R > 0$ be a desired transmission rate. For a fixed distance $d$ between the source and the relay, and a fixed transmission scheme $\mathcal{S}$, we define the coverage region as,

$$\mathcal{G}_\mathcal{S}(d) \triangleq \{\mathbf{a}_3 \,:\, C_\mathcal{S}(d, \mathbf{a}_3) \geq R\}$$

The concept of coverage is closely related to *outage* - we fix a target rate, and seek to maximize the geographic region, outside which an outage occurs.

## III. COMPARISON OF CF AND DF

In this section, we consider the coverage region when using the DF and CF approaches, for a fixed, given location of the relay $\mathbf{a}_2$. For reference, we also consider the *no-relay* (NR) coverage region, i.e, the coverage region when the relay is not used.

We would expect different locations of the destination $\mathbf{a}_3$ to favor different schemes. That is, for a fixed rate $R$, some locations would be covered only by CF, others only by DF and yet others by both. Surprisingly, however, the following result indicates a monotonic ordering of the coverage regions.

*Theorem 1:* Assume the channel model of Sec. II and let $R > 0$. Let $d_c$ be defined as follows,

$$d_c = \left( \frac{P_1}{2^R - 1} \right)^{1/\alpha} \quad (1)$$

1) If $d \leq d_c$, then

$$\mathcal{G}_{\text{DF}}(d) \supseteq \mathcal{G}_{\text{CF}}(d) \supseteq \mathcal{G}_{\text{NR}}(d)$$

2) If $d > d_c$, then

$$\mathcal{G}_{\text{CF}}(d) \supseteq \mathcal{G}_{\text{NR}}(d) \supseteq \mathcal{G}_{\text{DF}}(d) = \varnothing$$

The proof is provided in Appendix B.

We now interpret the results of Theorem 1. The condition $d \leq d_c$ determines whether the relay is still able to decode the information that is transmitted to it by the source. Whenever the relay is able to decode, we see that DF is the method of choice. DF is uniformly superior in transmission to any point, in the sense that it provides coverage at any point served by CF. However, if the relay is not able to decode the data, then DF cannot be applied, and $\mathcal{G}_{\text{DF}}(d) = \varnothing$. For such values of $d$, the best approach is CF, where the relay and destination combine their channel observations and perform collaborative decoding.

From a design perspective, while for $d \leq d_c$ DF enjoys a larger coverage region than CF, it suffers from a sharp drop in performance when $d$ crosses $d_c$. In practical settings, when the path loss exponent $\alpha$ is not known (and consequently $d_c$, as defined by (1), is not known), this may become an important disadvantage. CF, in comparison, enjoys a more graceful degradation with respect to changes in $d$.

The above results can equivalently be stated as follows: Consider the combined transmission strategy $\text{CF} \vee \text{DF}$, under which the various terminals are free to select the best of CF and DF. Theorem 1 implies that,

$$\mathcal{G}_{\text{CF} \vee \text{DF}}(d) = \begin{cases} \mathcal{G}_{\text{DF}}(d), & d \leq d_c; \\ \mathcal{G}_{\text{CF}}(d), & d > d_c; \end{cases}$$

Figs. 3 and 4 present numerical examples of the regions considered in Theorem 1. In Fig. 3, $d \leq d_c$ and in Fig. 4, $d > d_c$. In both figures we also compare the coverage regions with a region computed according to the upper-bound (UB) as provided by [5].

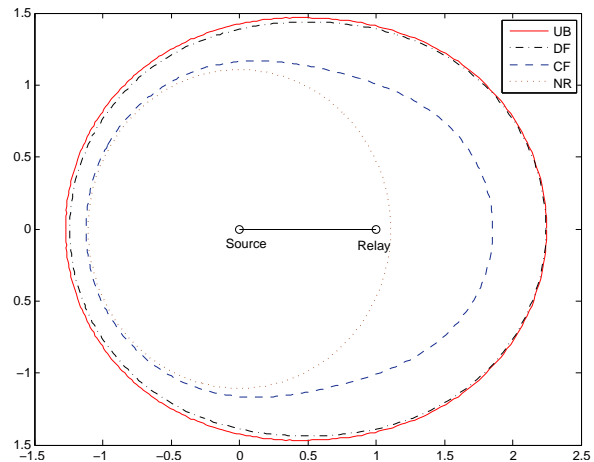

Fig. 3. Coverage regions when $d \leq d_c$, $P_1 = 10, P_2 = 10, \alpha = 3.52, R = 3$

## IV. COMPARISON WITH UNEQUAL RELAY PLACEMENT

The analysis of Sec. III focused on coverage regions for a given distance $d$ between the source and the relay. However, different schemes may favor different locations of the relay. Thus, in this section, we provide some interesting results that consider CF and DF when they are allowed *different* locations of the relay.

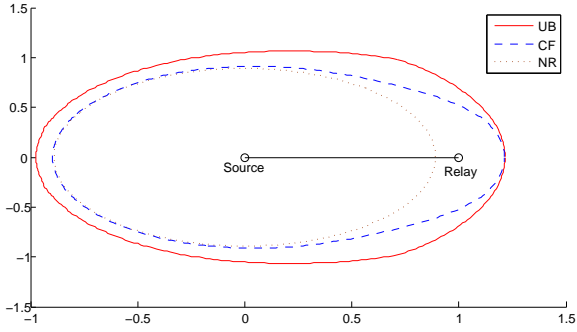

Fig. 4. Coverage regions when $d > d_c$, $P_1 = 10$, $P_2 = 10$, $\alpha = 3.52$, $R = 4$

*Theorem 2:* Let $R > 0$. Let $d_c$ be defined as in Theorem 1. Then the following assertions holds:

1) Assume $\alpha = 2$. Then there exists a non-negative positive fraction $0 < \gamma < 1/9$, independent of $R$, such that if $P_2 > \gamma \cdot P_1$, then for all $d > d_c$,

$$\mathcal{G}_{\mathrm{CF}}(d) \subseteq \mathcal{G}_{\mathrm{DF}}(d_c)$$

2) There exists $\beta(R) > 0$ such that if $P_2 < \beta(R) \cdot P_1$, then there exists $d_0$ such that,

$$\mathcal{G}_{\mathrm{CF}}(d_0) \not\subseteq \bigcup_{d>0} \mathcal{G}_{\mathrm{DF}}(d)$$

The proof of this theorem is provided in Appendix C

Theorem 2, combined with Theorem 1, provide us with insight into the choice between strategies DF and CF when the relay location may be optimized independently for each scheme. In the case of $\alpha = 2$, when the relay power is sufficiently large, DF is the method of choice regardless of the desired $R$. If we choose to place the relay at location $d \leq d_c$, then Theorem 1 tells us that $\mathcal{G}_{\mathrm{CF}}(d) \subseteq \mathcal{G}_{\mathrm{DF}}(d)$, and thus DF is superior. If $d > d_c$, then Theorem 2 tells us that $\mathcal{G}_{\mathrm{CF}}(d) \subseteq \mathcal{G}_{\mathrm{DF}}(d_c)$, and thus switching from CF and DF and repositioning the relay at $d = d_c$ would render superior performance.

Regardless of $\alpha$, if the power at the relay is sufficiently *low*, Theorem 2 tells us that there are locations and rates that may only be supported by CF. These cannot be supported by DF, regardless of where we place the relay.

## V. BOUNDS ON THE DF COVERAGE AREA

So far, our discussion has focused on a comparison of coverage regions with DF and CF. However, a natural question that arises is what the *area* (in normalized units of area) of the coverage region is, and the effect of the distance $d$ on it. In this section we partially answer this question for the DF coverage region in a few specific cases. We let $|\mathcal{G}_{\mathrm{DF}}(d)|$ denote this area, and begin with the following theorem.

*Theorem 3:* Assume $P_1 = P_2$ and $\alpha = 2$. Then for all $0 < d \leq d_c$,

$$\pi\sqrt{\lambda\gamma} \cdot d^2 \leq |\mathcal{G}_{\mathrm{DF}}(d)| \leq \pi\sqrt{\lambda\gamma} \cdot \frac{1 - a/2}{\sqrt{1-a}} \cdot d^2 \quad (2)$$

where,

$$\rho = \sqrt{1 - \left(\frac{d}{d_c}\right)^2}, \quad (3)$$

$$\lambda = \frac{1}{4} + \frac{1}{1-\rho} + \frac{\sqrt{2(1+\rho)}}{1-\rho^2} \quad (4)$$

$$\gamma = \frac{2}{1-\rho} - \frac{1}{4} \quad (5)$$

$$a = 1 - \frac{\gamma}{\lambda} \quad (6)$$

*Remark 1:*

1) Recall, from Theorem 3, that when $d > d_c$, $\mathcal{G}_{\mathrm{DF}}(d) = \varnothing$ and thus $|\mathcal{G}_{\mathrm{DF}}(d)| = 0$.
2) Observe that $\rho$ is is a function of $d$, and implicitly a function of of $R$ and $P_1$ through its dependence on $d_c$, which was defined by (1). Consequently, the other parameters are functions of these values too.

The proof of this theorem is achieved by bounding the coverage region from within by an ellipse whose boundary is given by the points $(\frac{d}{2} + \sqrt{\lambda}d\cos(\theta), \sqrt{\gamma}d\sin(\theta))$, and from outside by a conic whose boundary is given by the points $(\frac{d}{2} + \sqrt{\lambda}d\cos(\theta)\sqrt{1 - a\sin^2(\theta)}, \sqrt{\gamma}d\sin(\theta)\sqrt{1 - a\sin^2(\theta)}/\sqrt{1-a})$. These two bounding shapes, for the case of $P_1 = P_2 = 10, R = 1$, are plotted in Fig. 5, along with the true region (computed numerically). The details of the proof are provided in Appendix D.

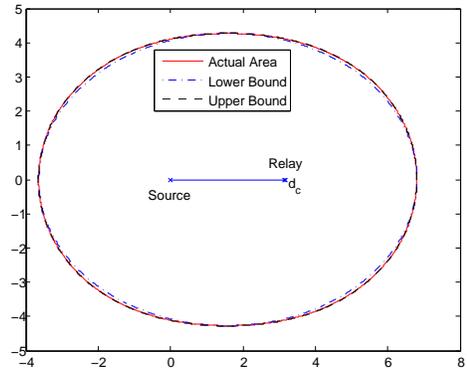

Fig. 5. Bounds for $\mathcal{G}_{\mathrm{DF}}$, $P_1 = P_2 = 10$, $R = 1$, $\alpha = 2$

Fig. 6 presents the bounds in (2), along with the true area (computed numerically) corresponding to the same parameters as Fig. 5. Examining this figure, we see that the bounds are very tight. By (2), the ratio between the two bounds is $(1 - a/2)/(\sqrt{1-a})$, which increases with $d$ from 1 to $\frac{6+2\sqrt{2}}{\sqrt{7(5+4\sqrt{2})}} \approx 1.02216$. Thus, the gap between the two bounds never exceeds $2.22\%$.

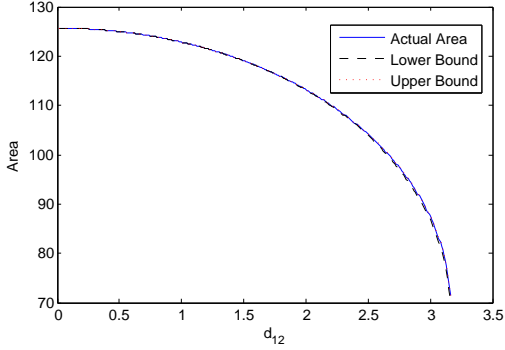

Fig. 6. Bounds for $\mathcal{G}_{\mathrm{DF}}$, $P_1 = P_2 = 10$, $R = 1$, $\alpha = 2$

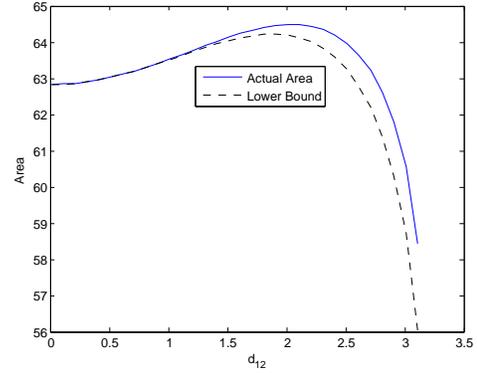

Fig. 7. Lower bound for $\mathcal{G}_{\mathrm{DF}}$, $P_1 = P_2 = 100$, $R = 1$, $\alpha = 4$

Examining Fig. 6, we may observe that $|\mathcal{G}_{\mathrm{DF}}(d)|$ approaches its maximum as $d \to 0$. To see this, observe that both bounds coincide as $d \to 0$, and the upper bound decreases with $d$. Thus, from the point of view of maximizing coverage, the relay should optimally be placed as close to the base station as possible. While this results holds for $\alpha = 2$, it is not generally true, as the discussion below will show.

The following theorem extends the lower bound of (2) to the case of $\alpha = 4$.

*Theorem 4:* Assume $P_1 = P_2$ and $\alpha = 4$. Then for all $0 < d \leq d_c$,

$$|\mathcal{G}_{\mathrm{DF}}(d)| \geq \pi \sqrt{\lambda \gamma} \cdot d^2 \tag{7}$$

where,

$$\rho = \sqrt{1 - \left(\frac{d}{d_c}\right)^4} \tag{8}$$

$$\gamma = \sqrt{\frac{2}{1-\rho} - 1/4} \tag{9}$$

and $\lambda$ is the largest real-valued solution of the equation,

$$\left(x - \frac{1}{4}\right)^4 - \frac{2}{1-\rho}\left(x - \frac{1}{4}\right)^2 - \frac{4}{1-\rho^2}\left(x - \frac{1}{4}\right) - \frac{1}{1-\rho^2} = 0 \tag{10}$$

This solution can be found analytically by applying the Ferrari method, see e.g. [6][page 32].
The proof of this theorem follows along the lines of the proof of the lower bound in Theorem 3, namely by bounding the coverage area from within by an ellipse. The details of the proof are provided in Appendix E.

Fig. 7 compares the lower bound of Theorem 4 with the true area (computed numerically), when $P_1 = P_2 = 100$ and $R = 1$.

*Lemma 1:* The lower bound (7) becomes tight as $d \to 0$.
The proof of this lemma is provided in Appendix F.

Recall that with $\alpha = 2$, we proved that the maximum coverage area was achieved when $d \to 0$. Lemma 1 enables us to show that this is *not* the case with $\alpha = 4$. To prove this, all we need to do is find a nonzero value of $d$ at which the lower bound is greater than its value at $d = 0$. Changing variables from $d$ to $\rho$ as in the proof of Lemma 1 (Appendix F), we obtain that at $\rho = .93$,

$$\begin{aligned}
|\mathcal{G}_{\mathrm{DF}}|_{\rho=.93} &\geq \pi\sqrt{\lambda\gamma}d^2_{\rho=.93} \\
&= 2.0441..\pi d_c^2 \\
&> 2\pi d_c^2 \\
&\stackrel{(a)}{=} |\mathcal{G}_{\mathrm{DF}}(d=0)|
\end{aligned}$$

where (a) follows as in the proof of Lemma 1.

## VI. ADDITIONAL CHANNEL MODELS

### A. Half Duplex Relay

The half-duplex relay model is characterized by a relay that cannot transmit and receive at the same time. A comprehensive discussion of this case is provided in [9], [5] and [10]. In this setting, we assume that there exists a positive fraction $t \in [0, 1]$ such that the relay is listening (receiving) during a proportion $t$ of the time, and transmitting during a proportion $1-t$ of the time. For this setting, we have the following theorem,

*Theorem 5:* Assume the half-duplex channel model and let $R > 0$. The following results hold.

1) If $t = 1/2$ and $d < d'_c$, where

$$d'_c = \left(\frac{P_1}{2^{2R} - 1}\right)^{1/\alpha} \tag{11}$$

then

$$\mathcal{G}_{\mathrm{DF}}(d) \supseteq \mathcal{G}_{\mathrm{CF}}(d) \supseteq \mathcal{G}_{\mathrm{NR}}(d)$$

2) If $d > d_c$, then for all $t \in [0, 1]$

$$\mathcal{G}_{\mathrm{CF}}(d) \supseteq \mathcal{G}_{\mathrm{NR}}(d) \supseteq \mathcal{G}_{\mathrm{DF}}(d)$$

The proof is provided in Appendix G. This theorem is weaker than Theorem 1, because $d'_c < d_c$. Furthermore, note that we have confined our attention to the case that $t = 1/2$.

## B. Random Fading Models

In this section, we modify the channel model of Sec. II to introduce some random fading. The new channel equations are now given by,

$$y_2[i] = \frac{h_{12}}{d_{12}^{\alpha/2}} e^{j\varphi_{12}[i]} x_1[i] + z_1[i]$$

$$y_3[i] = \frac{h_{13}}{d_{13}^{\alpha/2}} e^{j\varphi_{13}[i]} x_1[i] + \frac{h_{23}}{d_{23}^{\alpha/2}} e^{j\varphi_{23}[i]} x_2[i] + z_2[i]$$

We consider two fading models,

1) **Phase fading** $h_{kl} = 1$ for all $\{k,l\}$. $\varphi_{kl}[i]$ are uniformly distributed over $[0, 2\pi)$, and are jointly independent of one other, of the transmitted signals and the noise. Their time realizations are also independent.
2) **Rayleigh fading.** $\varphi_{kl}[i]$ is defined as in the phase fading case. $h_{kl}$ are Rayleigh distributed with parameter 1, independent, and remain fixed for the duration of the transmission.

In all cases, we assume that the realizations of the random variables $h_{kl}$ and $\varphi_{kl}[i]$ are known to the receivers but not to the transmitters.

We begin by considering the phase-fading model. In Appendix H-A we will show that Theorem 1 carries over directly to this case. Furthermore, Kramer *et al.* [5] have computed an upper bound (UB) on the capacity in the phase-fading model. In Appendix H-B we prove the following lemma.

*Lemma 2:* Assume $d \leq d_c$ where $d_c$ is given by (1). Then $\mathcal{G}_{\text{UB}}(d) = \mathcal{G}_{\text{DF}}(d)$.

We now proceed to provide some interesting observation for the Rayleigh fading model. This channel is no longer an ergodic channel, and thus we redefine the coverage region in terms of *outage probabilities* ([7], [5]).

*Definition 3:* Let $R > 0$ be a desired transmission rate and $0 < \epsilon < 1$ be the maximum tolerable outage probability. For a fixed distance $d$ between the source and the relay, and a fixed transmission scheme $\mathcal{S}$, we define the coverage region as,

$$\mathcal{G}_\mathcal{S}(d) \triangleq \{\mathbf{a}_3 \ : \ Pr[C_\mathcal{S}(d, \mathbf{a}_3) \geq R] \geq 1 - \epsilon\}$$

We define $\hat{d}_c$ in a manner analogous to $d_c$ of Theorem 1,

$$\hat{d}_c = \left( \frac{-P_1 \ln(1-\epsilon)}{2^R - 1} \right)^{1/\alpha}$$

$\hat{d}_c$ is the distance after which the relay cannot decode with probability greater than $1 - \epsilon$. At relay locations satisfying $d > \hat{d}_c$, DF cannot be applied.

Expressions for the achievable rates with CF and DF in this case are available in [5]. In Fig. 8 we have plotted the coverage regions for the various schemes in the Rayleigh fading model. Interestingly, the monotonic ordering of $\mathcal{G}_{\text{CF}}(d)$ and $\mathcal{G}_{\text{DF}}(d)$, which was observed in Theorem 1 for the simple (non-fading) model, is not maintained. Neither of the two regions contains the other.

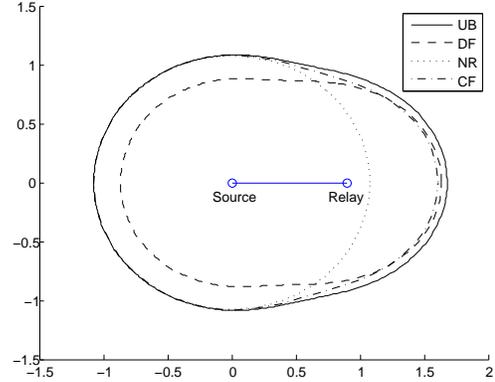

Fig. 8. Coverage regions with Rayleigh Fading and $d < \hat{d}_c, P_1 = 3, P_2 = 0.3, \alpha = 3.52, R = 1, \epsilon = 0.35, d = 0.9$

## VII. CONCLUSION

In this paper, we have introduced a new perspective on the relay channel, switching from maximizing rate at a fixed given destination location, to maximizing coverage for a given rate. This perspective opens up an array of new possibilities for research.

Our main contributions are first, the formulation of the problem in Sec. II. In Theorem 1, we have obtained the surprising result that for any given placement of the relay, one of the two common strategies (CF and DF) is uniformly optimal, and thus a relay that switches between the two is not required. Theorem 2 extends the comparison for the case of $\alpha = 2$ by allowing each of the strategies its own preferred relay location. In this case, it is interesting that the results depend on the power constraint on the relay. Theorems 1 and 2 imply that while DF often provides a larger coverage region than CF, it is also more sensitive to changes in the location of the relay and the path loss exponent $\alpha$. In contrast, CF is more robust and provides a more graceful degradation with respect to such changes. A natural question that arises, when considering coverage regions, is the numeric area of the region, as a function of the distance between the source and the relay. In Sec. V we have provided bounds in two special cases.

In Sec.VI, we have have extended our discussion to half-duplex and random-fading channel models. In particular, our results for fading channels indicate that the monotonic ordering of the DF and CF coverage regions no longer holds. A further study of the effect of relocating the relay, in such settings, is of great practical interest.

In this paper, we have focused on a one-relay setting. An interesting extension would consider two or more relays. In this case, the number of degrees of freedom grows substantially, as not only distance to the relays but also the angles between the line segments connecting them to the source, may be optimized. An analysis of the benefit from cooperation between the relays, is also of great interest.

## APPENDIX A
## GENERAL EXPRESSIONS AND NOTATION

### A. DF and CF Achievable Rates

The achievable rates, with DF and CF, that were computed by [5], [8],

$$C_{\mathrm{DF}} = \max_{0 \leq \rho \leq 1} \min \left\{ \log\left(1 + \frac{P_1}{d_{12}^\alpha}(1-\rho^2)\right), \right.$$
$$\left. \log\left(1 + \frac{P_1}{d_{13}^\alpha} + \frac{P_2}{d_{23}^\alpha} + \frac{2\rho\sqrt{P_1 P_2}}{d_{13}^{\alpha/2} d_{23}^{\alpha/2}}\right) \right\} \quad (12)$$

$$C_{\mathrm{CF}} = \log\left(1 + \frac{P_1}{d_{12}^\alpha(1+\hat{N}_2)} + \frac{P_1}{d_{13}^\alpha}\right) \quad (13)$$

where $\hat{N}_2$ is given by,

$$\hat{N}_2 = \frac{P_1(1/d_{12}^\alpha + 1/d_{13}^\alpha) + 1}{P_2/d_{23}^\alpha} \quad (14)$$

### B. Switch to Polar Coordinates

We frequently use the polar coordinates $x$ and $\theta$ to parameterize the destination's location, i.e. we assume the destination is placed at,

$$\mathbf{a}_3 = (x\cos\theta, x\sin\theta) \quad (15)$$

Letting $d$ denote the distance between the source and the relay, the distances $d_{12}, d_{13}$ and $d_{23}$ (see Sec. II-A) now satisfy, $d_{12} = d$, $d_{13} = x$, and $d_{23}^2 = d^2 + x^2 - 2dx\cos\theta$.

## APPENDIX B
## PROOF OF THEOREM 1

We begin by noting that for all $d > 0$, the result $\mathcal{G}_{\mathrm{CF}}(d) \supseteq \mathcal{G}_{\mathrm{NR}}(d)$ is straightforward from (13) and from the observation that,

$$C_{\mathrm{NR}} = \log\left(1 + \frac{P_1}{d_{13}^\alpha}\right) \quad (16)$$

We now prove Part 2 of the theorem, which is the easier of the two parts. Whenever $d > d_c$, by (1),

$$R > \log\left(1 + \frac{P_1}{d^\alpha}\right) = \max_{\rho \in [0,1]} \log\left(1 + \frac{P_1}{d^\alpha}(1-\rho^2)\right) \geqslant C_{\mathrm{DF}}$$

where the last inequality was obtained by (12), recalling that $d_{12} = d$. This is true regardless of the destination location $\mathbf{a}_3$, and thus DF cannot support a rate of $R$ anywhere and $\mathcal{G}_{\mathrm{DF}}(d) = \phi$.

Before proceeding to the proof of Part 1 of the theorem, we introduce the following notation. For a given scheme $\mathcal{S}$ ($\mathcal{S}$ being DF or CF), and a given $\theta$, we define $x_\mathcal{S}(\theta)$ as the maximum $x$ such that a destination with polar coordinates $(x, \theta)$ (see Appendix A) is contained in $\mathcal{G}_\mathcal{S}(d)$. The proof now focuses on showing that $x_{\mathrm{DF}}(\theta) \geq x_{\mathrm{CF}}(\theta)$ for all $\theta$.

By (12), we may obtain a lower bound on $C_{\mathrm{DF}}$ by restricting the maximization to $\rho = 0$. Thus,

$$C_{\mathrm{DF}} \geq \min\left\{\log\left(1+\frac{P_1}{d_{12}^\alpha}\right), \log\left(1+\frac{P_1}{d_{13}^\alpha}+\frac{P_2}{d_{23}^\alpha}\right)\right\} \quad (17)$$

The following lemma examines this expression at $x = x_{\mathrm{DF}}(\theta)$.

*Lemma 3:* For all $\theta$, consider the point $\mathbf{a}_3$ whose polar coordinates are given by $(\theta, x_{\mathrm{DF}}(\theta))$. Then the following holds at $\mathbf{a}_3$,

$$\frac{P_1}{d_{12}^\alpha} \geqslant \frac{P_1}{d_{13}^\alpha} + \frac{P_2}{d_{23}^\alpha} \quad (18)$$

*Proof:* Assume, by contradiction, that (18) does not hold. Letting $x = x_{\mathrm{DF}}(\theta)$, we will now show that we may increase $x$ and preserve $C_{\mathrm{DF}} \geq R$, contradicting the maximality of $x_{\mathrm{DF}}(\theta)$.

Consider (12). Whenever (18) does not hold, we have, for all $\rho$,

$$\log\left(1 + \frac{P_1}{d_{12}^\alpha}(1-\rho^2)\right) \leq \log\left(1 + \frac{P_1}{d_{12}^\alpha}\right)$$
$$< \log\left(1 + \frac{P_1}{d_{13}^\alpha} + \frac{P_2}{d_{23}^\alpha} + \frac{2\rho\sqrt{P_1 P_2}}{d_{13}^{\alpha/2} d_{23}^{\alpha/2}}\right)$$

Thus, the minimization in (12) is achieved by the first term. Changing $x$ does not affect this term. Increasing $x$ affects $d_{13}$ and $d_{23}$ and thus reduces the second term. However, by a continuity argument, we may increase $x$ slightly without altering the invalidity of (12). Thus, the minimization in (12) would still be achieved by the first term, and $C_{\mathrm{DF}}$ will not change. Therefore, if $C_{\mathrm{DF}} \geq R$ at $x = x_{\mathrm{DF}}(\theta)$, it will remain so after we increase $x$ a little, producing the desired contradiction. ∎

By Lemma 3, (17) implies that at $x = x_{\mathrm{DF}}(\theta)$,

$$C_{\mathrm{DF}} \geq \log\left(1 + \frac{P_1}{d_{13}^\alpha} + \frac{P_2}{d_{23}^\alpha}\right) \quad (19)$$
$$> C_{\mathrm{CF}} \quad (20)$$

The last inequality may be obtained by simple arithmetic using (13) and (14). By a continuity argument, $C_{\mathrm{DF}} = R$ at $(\theta, x_{\mathrm{DF}}(\theta))$, and thus the above inequality implies that $C_{\mathrm{CF}} < R$ at this point.

We now argue that this result implies that $\mathcal{G}_{\mathrm{DF}}(d) \supseteq \mathcal{G}_{\mathrm{CF}}(d)$. This would follow if we could show that $C_{\mathrm{CF}}$ decreases in the range $x > x_{\mathrm{DF}}(\theta)$, for all $\theta$.

In the range $x > d$, $d_{13}, d_{23}$ can easily be shown to be increasing functions. This implies that $\hat{N}_2$ increases (see (14)), and consequently $C_{\mathrm{CF}}$ decreases (see (13)). Thus, our desired result would now follow if we could prove that $x_{\mathrm{DF}}(\theta) > d$ for all $\theta$. To see this, observe that by (16) and (1), $\mathcal{G}_{\mathrm{NR}}$ is a sphere with radius $d_c$. We have shown that $\mathcal{G}_{\mathrm{NR}} \subset \mathcal{G}_{\mathrm{CF}}$. Thus, if $C_{\mathrm{CF}} < R$ at $(\theta, x_{\mathrm{DF}}(\theta))$ (as we have shown above), then $x_{\mathrm{DF}}(\theta) \geq d_c \geq d$ (the last inequality being one of the conditions of this part of the theorem). This is precisely the result we sought, thus completing the proof of Part 1 of the theorem. ∎

## APPENDIX C
## PROOF OF THEOREM 2

Before beginning with the proof, we argue that we may assume, without loss of generality, that $P_1$ (the power of the source) is 1. To see this, observe that if $P_1 \neq 1$, we may replace $P_1$ and $P_2$ by $\hat{P}_1 = 1$ and $\hat{P}_2 = P_2/P_1$. This substitution preserves the ratio $\hat{P}_1/\hat{P}_2 = P_1/P_2$, and has the effect of scaling the DF and CF coverage regions. That is, if we relocate the relay and the destination from any locations $\mathbf{a}_2$ and $\mathbf{a}_3$ (respectively) to $1/P_1^{1/\alpha} \cdot \mathbf{a}_2$ and $1/P_1^{1/\alpha} \cdot \mathbf{a}_3$, the achievable rates $C_{\mathrm{CF}}$ and $C_{\mathrm{DF}}$ would remain unchanged.

Under these assumptions, given that the relay is placed at $(d,0)$, and using the notation of Sec. A, the expressions for the DF and CF achievable rates (12) and (13) may be rewritten as,

$$C_{\mathrm{DF}} = \max_{0 \leq \rho \leq 1} \min \left\{ \log\left(1 + \frac{1}{d^\alpha}(1-\rho^2)\right), \right.$$
$$\left. \log\left(1 + \frac{1}{x^\alpha} + \frac{P_2}{d_{23}^\alpha} + \frac{2\rho\sqrt{P_2}}{d^{\alpha/2} d_{23}^{\alpha/2}}\right) \right\} \quad (21)$$

$$C_{\mathrm{CF}} = \log\left(1 + \frac{1}{d^\alpha(1+\hat{N}_2)} + \frac{1}{x^\alpha}\right) \quad (22)$$

and (14) by,

$$\hat{N}_2 = \frac{1/d^\alpha + 1/x^\alpha + 1}{P_2/d_{23}^\alpha} \quad (23)$$

### A. Proof of Part 1 of Theorem 2

In this section, we assume that $\alpha = 2$. Rather than focus on $C_{\mathrm{DF}}$ and $C_{\mathrm{CF}}$ as in (21) and (22), the following lemma allows us to revert to two alternative functions.

*Lemma 4:* Let $C_{\mathrm{DF}}(x,\theta,d)$ and $C_{\mathrm{CF}}(x,\theta,d)$ denote the achievable rates when the relay is placed at $(d,0)$ and the destination location is derived from $x$ and $\theta$. Let,

$$\tilde{C}_{\mathrm{DF}}(x,\theta) \triangleq \log\left(1 + \frac{1}{x^2} + \frac{P_2}{d_{23,c}^2}\right) \quad (24)$$

$$C_{\mathrm{CF}}^+(x,\theta) \triangleq \log\left(1 + \frac{1}{x^2} + \frac{P_2}{\inf_{d \geq d_c}(P_2 d^2 + d_{23}^2)}\right) \quad (25)$$

where $d_{23,c}$ is the distance from a relay placed at $(d_c, 0)$ to the destination, and $d_{23}$ is the same for a relay placed at $(d,0)$. Then the following holds:

1) $\tilde{C}_{\mathrm{DF}}(x,\theta) \geq R$ if and only if $C_{\mathrm{DF}}(x,\theta,d_c) \geq R$. Equivalently,
$$\mathcal{G}_{\mathrm{DF}}(d_c) = \left\{(x\cos\theta, x\sin\theta) : \tilde{C}_{\mathrm{DF}}(x,\theta) \geq R\right\}$$

2) $C_{\mathrm{CF}}^+(x,\theta)$ upper bounds $C_{\mathrm{CF}}(x,\theta,d)$ for all $d \geq d_c$, and therefore,
$$\bigcup_{d \geq d_c} \mathcal{G}_{\mathrm{CF}}(d) \subset \left\{(x\cos\theta, x\sin\theta) : C_{\mathrm{CF}}^+(x,\theta) \geq R\right\}$$

*Proof:* Part 1 follows from the observation that when $d = d_c$, by (1), any choice of $\rho$ in (21) other than $\rho = 0$ would render $C_{\mathrm{DF}} < R$.

Part 2 follows by bounding the term $d^2(1+\hat{N}_2)$, which appears in (22),

$$d^2(1+\hat{N}_2) = \frac{d^2(1/d^2 + 1/x^\alpha + 1 + P_2/d_{23}^2)}{P_2/d_{23}^2}$$
$$\geq \frac{d^2 d_{23}^2(1/d^2 + P_2/d_{23}^2)}{P_2}$$
$$= \frac{d_{23}^2 + P_2 d^2}{P_2}$$

∎

Our aim now is to show that wherever $C_{\mathrm{CF}}^+(x,\theta) \geq R$, also $\tilde{C}_{\mathrm{DF}}(x,\theta) \geq R$. In the sequel, we use the shorthand notation $s \triangleq \cos\theta$.

*Lemma 5:* Let $\theta$ be fixed,
1) If $s \leq 0$, then for all $x > 0$, $\tilde{C}_{\mathrm{DF}}(x,\theta) \geq C_{\mathrm{CF}}^+(x,\theta)$.
2) Assume $s > 0$ and let $x_1$ be given by,
$$x_1 = \frac{1}{s} \frac{\sqrt{P_2+1}}{\sqrt{P_2+1} - \sqrt{P_2}} \cdot d_c \quad (26)$$

Then the following holds,
$$C_{\mathrm{CF}}^+(x,\theta) \leq \tilde{C}_{\mathrm{DF}}(\min(x,x_1),\theta)$$

*Proof:* Observe, by (24) and (25) that $\tilde{C}_{\mathrm{DF}}(x,\theta) \geq C_{\mathrm{CF}}^+(x,\theta)$ if and only if $f_{\min} \geq d_{23,c}^2$ where $f_{\min}$ is given by,
$$f_{\min} = \inf_{d \geq d_c}(P_2 d^2 + d_{23}^2) \quad (27)$$

Let $f(d)$ be defined by,
$$f(d) \triangleq P_2 d^2 + d_{23}^2$$
$$= (P_2+1)d^2 + x^2 - 2dxs \quad (28)$$

where the last equality was obtained by the cosine theorem. Taking the derivative of $f(d)$, we obtain that the unconstrained minimum (i.e., the minimum in the range $d \in (-\infty, \infty)$) is obtained at,
$$d_{\min} = \frac{xs}{P_2+1} \quad (29)$$

If $s \leq 0$ then $d_{\min} \leq 0 < d_c$, and thus $f_{\min}$, which is the minimum in the range $d \geq d_c$, is obtained at $d_c$. Thus,
$$f_{\min} = f(d_c) = P_2 d_c^2 + d_{23,c}^2 > d_{23,c}^2,$$

and our desired result of Part 1 of the lemma follows.

To prove Part 2, we begin by considering the range $x \in [0, x_1]$. We further divide the interval $x \in [0, x_1]$ into two overlapping intervals, $[0, x']$ and $[x'', x_1]$ and prove the desired result separately in each interval. We define,
$$x' = \frac{1}{s} \cdot d_c$$
$$x'' = \frac{1}{s} \frac{\sqrt{P_2+1}}{\sqrt{P_2+1} + \sqrt{P_2}} \cdot d_c$$

Clearly, $x'' < x' < x_1$, and thus the two intervals overlap.

If $x \leq x'$ then $d_{\min} < d_c$ (by (29)), and thus $f_{\min} \geq d_{23,c}^2$ as in the proof of Part 1 of the lemma. Thus, the desired result is obtained for $x \in [0, x']$.

We proceed to consider the interval $x \in [x'', x_1]$. The minimum $f_{\min}$ (constrained to $d \geq d_c$) clearly satisfies $f_{\min} \geq f(d_{\min})$. By (28) and (29), we have, after some algebraic manipulations,

$$f(d_{\min}) = x^2 \left(1 - \frac{s^2}{P_2 + 1}\right) \qquad (30)$$

Thus, a sufficient condition for $f_{\min} \geq d_{23,c}^2$ is,

$$x^2 \left(1 - \frac{s^2}{P_2 + 1}\right) \geq d_{23,c}^2 \qquad (31)$$

Applying the cosine theorem and rearranging the terms, we obtain the inequality,

$$\frac{s^2}{P_2 + 1} \cdot x^2 - (2 d_c s) x + d_c^2 \leq 0 \qquad (32)$$

The left hand side of this inequality is a parabola in $x$, whose roots are $x''$ and $x_1$. Thus, the inequality is satisfied for all $x \in [x'', x_1]$. Consequently, by the above discussion, $f_{\min} \geq d_{23,c}^2$ in this interval, as desired. This completes the proof on the lemma in the interval $x \in [0, x_1]$.

Before proceeding to the interval $[x_1, \infty)$, observe that $x_1$ satisfies (32) and consequently (31), with equality. Equivalently,

$$x_1^2 \left(1 - \frac{s^2}{P_2 + 1}\right) = d_{23,c}^2 \qquad (33)$$

We now consider $x \geq x_1$. Observe that the discussion leading to (30) remains valid in this range. By (25),

$$\begin{aligned}
C_{\mathrm{CF}}^+(x, \theta) &= \log\left(1 + \frac{1}{x^2} + \frac{P_2}{\inf_{d \geq d_c}(f(d,x,\theta))}\right) \\
&\leq \log\left(1 + \frac{1}{x^2} + \frac{P_2}{\inf_{d \geq 0} f(d,x,\theta)}\right) \\
&\stackrel{(a)}{=} \log\left(1 + \frac{1}{x^2} + \frac{P_2}{x^2\left(1 - \frac{s^2}{P_2+1}\right)}\right) \\
&\stackrel{(b)}{\leq} \log\left(1 + \frac{1}{x_1^2} + \frac{P_2}{x_1^2\left(1 - \frac{s^2}{P_2+1}\right)}\right) \\
&\stackrel{(c)}{=} \log\left(1 + \frac{1}{x_1^2} + \frac{P_2}{d_{23,c}^2}\right) \\
&= \tilde{C}_{\mathrm{DF}}(x_1, \theta)
\end{aligned}$$

where (a) is obtained by (30), (b) follows from $x \geq x_1$, and (c) is obtained by (33). ∎

Lemmas 4 and 5 imply that a sufficient condition for Part 1 of Theorem 2 is $\tilde{C}_{\mathrm{DF}}(x_1, \theta) \leq R$ for all $\theta$ such that $s = \cos\theta > 0$ (recall that $x_1$ is a function of $\theta$) and all $P_2 > 1/9$ (recall that we have assumed that $P_1 = 1$).

Using (1) and (24), $\tilde{C}_{\mathrm{DF}}(x_1, \theta) \leq R$ can be rewritten as,

$$\log\left(1 + \frac{P_2}{d_{23,c}^2} + \frac{1}{x_1^2}\right) \leq \log\left(1 + \frac{1}{d_c^2}\right) \qquad (34)$$

We now introduce the notations,

$$A = \frac{1}{s} \frac{\sqrt{P_2 + 1}}{\sqrt{P_2 + 1} - \sqrt{P_2}}, \quad B = \left(1 - \frac{s^2}{P_2 + 1}\right) \qquad (35)$$

Observe that $A$ and $B$ are functions of $s$. With these notations, using (26) and (33), we obtain, $x_1 = A \cdot d_c$ and $d_{23,c}^2 = B x_1^2 = B A^2 \cdot d_c^2$. Plugging these identities into (34), with some algebraic manipulations, leads to the inequality,

$$P_2 \leq B(A^2 - 1) \qquad (36)$$

The right hand side of the above inequality is a descending function of $s$ in the range $s \in [0, 1]$. Thus, a necessary and sufficient condition for the inequality to be valid for all $s \in [0, 1]$ is that the inequalities hold for $s = 1$. We thus confine our attention to this case. Using (35), we may rewrite (36) as,

$$P_2 \leq \left(1 - \frac{1}{P_2 + 1}\right)\left[\left(\frac{\sqrt{P_2+1}}{\sqrt{P_2+1} - \sqrt{P_2}}\right)^2 - 1\right]$$

We now proceed to develop this inequality,

$$\begin{aligned}
P_2 &\leq \frac{P_2}{P_2+1}\left(\frac{P_2+1}{(\sqrt{P_2+1} - \sqrt{P_2})^2} - 1\right) \\
(\sqrt{P_2+1} - \sqrt{P_2})^2 &\stackrel{(a)}{\leq} \frac{P_2+1}{P_2+2} \\
\sqrt{P_2+1} - \sqrt{P_2} &\stackrel{(b)}{\leq} \frac{\sqrt{P_2+1}}{\sqrt{P_2+2}} \\
\sqrt{(P_2+1)(P_2+2)} &\stackrel{(c)}{\leq} \sqrt{P_2(P_2+2)} + \sqrt{P_2+1} \\
(P_2+1)(P_2+2) &\stackrel{(d)}{\leq} P_2(P_2+2) + P_2 + 1 + \\
&\quad + 2\sqrt{P_2(P_2+2)(P_2+1)} \\
1 &\leq 2\sqrt{P_2(P_2+2)(P_2+1)} \\
\frac{1}{4} &\stackrel{(e)}{\leq} P_2(P_2+2)(P_2+1) \qquad (37)
\end{aligned}$$

(a) is obtained by multiplying both sides by $(P_2+1)/P_2 \cdot (\sqrt{P_2+1} - \sqrt{P_2})^2$. (b) is obtained by taking the square root of both sides. (c) is obtained by multiplying both sides by $\sqrt{P_2+2}$ and rearranging terms. (d) is obtained by taking the squares of both sides. (e) is obtained by dividing by 2 and taking the square of both sides.

Finally, the right hand side of (37) is an ascending function of $P_2$. The inequality can be verified to be satisfied for $P_2 = 1/9$, and thus it is satisfied for all $P_2 > 1/9$. ∎

### B. Proof of Part 2 of Theorem 2

We begin with an outline of the proof. The proof follows from the observation that regardless of how low the power at the relay may be, if the destination is close enough to the relay, then arbitrarily high reliability may be achieved in the link between them. With CF, the relay and the destination cooperate and form a virtual antenna array. With DF, the relay must decode alone, and is thus confined to the no-relay coverage region (i.e., $d$ must not exceed $d_c$). If its power is sufficiently low, it cannot service a destination that is beyond the no-relay region.

We confine our attention in this section to destinations $\mathbf{a}_2$ of the form $(0, x)$ where $x > d_c$. We begin by examining $C_{\text{DF}}$. By (21) (assuming, as discussed above, that $P_1 = 1$),

$$\limsup_{P_2 \to 0} C_{\text{DF}} \leq \lim_{P_2 \to 0} \max_{0 \leq \rho \leq 1} \log\left(1 + \frac{1}{x^\alpha} + \frac{P_2}{d_{23}^\alpha} + \frac{2\rho\sqrt{P_2}}{x^{\alpha/2} d_{23}^{\alpha/2}}\right)$$

$$= \log\left(1 + \frac{1}{x^\alpha}\right) < \log\left(1 + \frac{1}{d_c^\alpha}\right) = R$$

The last equality follows by (1). Thus, at any $x > d_c$, for $P_2$ that is too small, $C_{\text{DF}} < R$, and thus we may not support the desired rate of $R$.

We now turn to examine $C_{\text{CF}}$. We wish to find a destination location $\mathbf{a}_3 = (0, x)$, where $x > d_c$, such that regardless of how low the relay power $P_2$ may be, by positioning the relay close enough to the destination, we will be able to support the desired rate of $R$. Combining this with our above result for DF will conclude the proof.

Let $x > d_c$ be arbitrary. We will examine relay locations given by $(0, d)$, where $d = x - P_2^{1/\alpha}$. With this choice, $P_2/d_{23}^\alpha = 1$.

$$\lim_{P_2 \to 0} C_{\text{CF}} = \lim_{P_2 \to 0} \log\left(1 + \frac{1}{d^\alpha(1 + \hat{N}_2)} + \frac{1}{x^\alpha}\right)$$

Where we have used $d_{12} = d$, and $d_{13} = x$. Now, it is straightforward to verify that the following holds, recalling that $d$ is now a function of $P_2$, and it satisfies $P_2/d_{23}^\alpha = 1$.

$$\lim_{P_2 \to 0} d^\alpha (1 + \hat{N}_2) = 2(1 + x^\alpha)$$

Thus,

$$\lim_{P_2 \to 0} C_{\text{CF}} = \log\left(1 + \frac{1}{2(1 + x^\alpha)} + \frac{1}{x^\alpha}\right)$$

Now, taking the limit as $x \to d_c$, we obtain:

$$\lim_{x \to d_c}\left(\lim_{P_2 \to 0} C_{\text{CF}}\right) = \log\left(1 + \frac{1}{2(1 + d_c^\alpha)} + \frac{1}{d_c^\alpha}\right)$$

$$> \log\left(1 + \frac{1}{d_c^\alpha}\right) = R$$

where the last equality follows by (1). Thus, there exists $x_0 > d_c$, such that for arbitrarily small $P_2$, the achievable rate with CF (when the relay is appropriately placed) is greater than $R$. This is exactly what we set out to prove, thus concluding the proof of the theorem. ∎

## APPENDIX D
## PROOF OF THEOREM 3

### A. The Lower Bound

As noted in Section V, the proof of the lower bound follows by bounding the coverage region from within by an ellipse whose boundary is given by the points,

$$\left(\frac{d}{2} + \sqrt{\lambda} d \cos(\theta), \sqrt{\gamma} d \sin(\theta)\right) \quad (38)$$

We let $\mathcal{A}$ denote this ellipse. We begin by focusing on the boundary of $\mathcal{A}$, and show that it lies within the DF coverage region. We will later show that this implies that interior of $\mathcal{A}$ is contained in the coverage region as well.

We must therefore show that at each point $\mathbf{a}_3$ on the boundary of $\mathcal{A}$,

$$C_{\text{DF}}(P_2 = P_1, \alpha = 2) \geq R \quad (39)$$

where $C_{\text{DF}}$ is given by (12).

Note that by (38), at a point on the boundary of $\mathcal{A}$, the following holds,

$$d_{13}^2 = d^2\left((\frac{1}{2} + \sqrt{\lambda}\cos(\theta))^2 + (\sqrt{\gamma}\sin(\theta))^2\right) \quad (40)$$

$$d_{23}^2 = d^2\left((\frac{1}{2} - \sqrt{\lambda}\cos(\theta))^2 + (\sqrt{\gamma}\sin(\theta))^2\right) \quad (41)$$

Consider (12). It is easy to observe that whenever $C_{\text{DF}}$ is greater or equal to $R$, then selecting $\rho$ to coincide with (3) would at worst reduce $C_{\text{DF}}$ to equal $R$. This is because this choice renders the first minimization term equal to $R$ and can only increase the second term. Thus, we assume, without loss of generality, that this value of $\rho$ maximizes (12).

Thus, proving (39) is equivalent to proving (recall that $P_1 = P_2$),

$$\log\left(1 + \frac{P_1}{d_{13}^2} + \frac{P_1}{d_{23}^2} + \frac{2\rho P_1}{d_{13} d_{23}}\right) \geq R = \log\left(1 + \frac{P_1}{d_{12}^2}(1 - \rho^2)\right).$$

Substituting (40) and (41) into the above equations, after some manipulations, we get (42) at the top of the page.

We now define $s \triangleq \sin^2(\theta)$. We also define,

$$t(s) \triangleq \frac{1}{4} + \lambda - (\lambda - \gamma)s \quad (43)$$

$$f(s) \triangleq t^2(s) - \lambda + \lambda s \quad (44)$$

Using this, equation (42) reduces to

$$\frac{2t(s)}{f(s)} + \frac{2\rho}{\sqrt{f(s)}} - (1 - \rho^2) \geq 0.$$

We will soon show (Lemma 6) that $f(s) > 0$, and thus, multiplying above equation by $-f(s)$, we obtain that it is enough to show that

$$h(s) \triangleq -2t(s) - 2\rho\sqrt{f(s)} + (1 - \rho^2)f(s) \leq 0,$$

for all $s \in [0, 1]$. We will soon show (Lemma 7) that $h(s)$ is zero on the end points of the interval $s \in [0, 1]$; hence it is enough to prove that $h(s)$ is convex. This will soon be proved in Lemma 8.

The above discussion implies that the boundary of $\mathcal{A}$ lies within $\mathcal{G}_{\text{DF}}$. We now argue that this implies that the interior of $\mathcal{A}$ is contained in $\mathcal{G}_{\text{DF}}$ as well. Consider a point $\mathbf{a}_3$ in the interior of $\mathcal{A}$, and consider its polar coordinates (15). If $x \leq d$, then it can easily be observed from (13) that $C_{\text{DF}} \geq R$ (recall that $d_{13} = x$, $d_{12} = d$ and that we have assumed $d \leq d_c$). Now, keeping $\theta$ fixed, consider the range $\{x : d < x < x_\mathcal{A}(\theta)\}$, where $x_\mathcal{A}(\theta)$ lies on the boundary of $\mathcal{A}$. Both $d_{13}$ and $d_{23}$ can easily be shown to be ascending functions of $x$ in this

$$\frac{1}{(\frac{1}{2} + \sqrt{\lambda}\cos(\theta))^2 + (\sqrt{\gamma}\sin(\theta))^2} + \frac{1}{(\frac{1}{2} - \sqrt{\lambda}\cos(\theta))^2 + (\sqrt{\gamma}\sin(\theta))^2}$$
$$+ \frac{2\rho}{\sqrt{((\frac{1}{2} + \sqrt{\lambda}\cos(\theta))^2 + (\sqrt{\gamma}\sin(\theta))^2)}\sqrt{((\frac{1}{2} - \sqrt{\lambda}\cos(\theta))^2 + (\sqrt{\gamma}\sin(\theta))^2)}} \geq 1 - \rho^2 \quad (42)$$

range, and thus by (12), $C_{\text{DF}}$ is a descending function of $x$. Therefore, if $C_{\text{DF}} \geq R$ at the boundary point $x_{\mathcal{A}}(\theta)$, then it is greater than $R$ at any point satisfying $x < x_{\mathcal{A}}(\theta)$.

We now prove some of the lemmas that were used in the above discussion.

*Lemma 6:*
1) $\lambda - \gamma \geq \frac{1}{2}$
2) $t'(s) < 0$ for all $s \in [0,1]$
3) $f(s)$ is convex
4) $f'(s) \leq 0$ for all $s \in [0,1]$
5) $f(s) \geq \frac{4}{(1-\rho)^2} > 0$ for all $s \in [0,1]$

*Proof:*

1)
$$\lambda - \gamma - \frac{1}{2} = \frac{1}{4} + \frac{1}{1-\rho} + \frac{\sqrt{2(1+\rho)}}{1-\rho^2} - \frac{2}{1-\rho} + \frac{1}{4} - \frac{1}{2}$$
$$= \frac{\sqrt{2(1+\rho)} - (1+\rho)}{1-\rho^2}$$
$$= \frac{\sqrt{1+\rho}}{1-\rho^2}(\sqrt{2} - \sqrt{1+\rho}) \geq 0$$

2) $t'(s) = -(\lambda - \gamma) < 0$ by Part (1) of this Lemma.
3) $f''(s) = 2t'^2(s) = 2(\lambda - \gamma)^2 > 0$ by Part (1) of this Lemma.
4) $f'(s) = 2t(s)(\gamma - \lambda) + \lambda$. As $t(s)$ is decreasing function (part (2) of this lemma), we need to prove that $2t(1)(\gamma - \lambda) + \lambda \leq 0$, or $2(\frac{1}{4} + \gamma)(\lambda - \gamma) \geq \lambda$. Or,

$$2\left(\frac{2}{1-\rho}\right)\left(\frac{1}{2} - \frac{1}{1-\rho} + \frac{\sqrt{2(1+\rho)}}{1-\rho^2}\right)$$
$$\geq \frac{1}{4} + \frac{1}{1-\rho} + \frac{\sqrt{2(1+\rho)}}{1-\rho^2}$$

After some manipulations, this is equivalent to proving
$$(3+\rho)\left(\sqrt{\frac{2}{1+\rho}} + \frac{1-\rho}{4}\right) \geq 4$$

which is true since the left hand side decreases with $\rho \in [0,1)$ and goes to 4 as $\rho \to 1$.

5) As $f'(s) \leq 0$ (by previous part of this Lemma), $f(s) \geq f(1) = t^2(1) = \frac{4}{(1-\rho)^2}$.

*Lemma 7:* $h(s) = 0$ on the end points of the interval $s \in [0,1]$, that is
1) $h(0) = 0$
2) $h(1) = 0$

*Proof:*

1)
$$h(0) = -2t(0) - 2\rho\sqrt{f(0)} + (1-\rho^2)f(0)$$
$$= -2\left(\frac{1}{4} + \lambda\right) - 2\rho\left(\lambda - \frac{1}{4}\right)$$
$$+ (1-\rho^2)\left(\lambda - \frac{1}{4}\right)^2$$
$$= -2\left(\frac{1}{2} + \frac{1}{1-\rho} + \frac{\sqrt{2(1+\rho)}}{1-\rho^2}\right)$$
$$- 2\rho\left(\frac{1}{1-\rho} + \frac{\sqrt{2(1+\rho)}}{1-\rho^2}\right)$$
$$+ (1-\rho^2)\left(\frac{1}{1-\rho} + \frac{\sqrt{2(1+\rho)}}{1-\rho^2}\right)^2$$
$$= -2\left(\frac{1}{2} + \frac{1+\rho}{1-\rho} + \frac{\sqrt{2(1+\rho)}}{1-\rho}\right)$$
$$+ \frac{1+\rho}{1-\rho}\left(1 + \frac{2}{1+\rho} + 2\frac{\sqrt{2(1+\rho)}}{1+\rho}\right)$$
$$= -1 - \frac{1+\rho}{1-\rho} + \frac{2}{1-\rho}$$
$$= 0$$

2)
$$h(1) = -2t(1) - 2\rho\sqrt{f(1)} + (1-\rho^2)f(1)$$
$$= -2\left(\frac{1}{4} + \gamma\right) - 2\rho\left(\frac{1}{4} + \gamma\right)$$
$$+ (1-\rho^2)\left(\frac{1}{4} + \gamma\right)^2$$
$$= (1/4 + \gamma)\left(-2 - 2\rho + (1-\rho^2)\left(\frac{2}{1-\rho}\right)\right)$$
$$= 0$$

∎

*Lemma 8:* $h(s)$ is convex.
*Proof:*
$$h(s) \triangleq -2t(s) - 2\rho\sqrt{f(s)} + (1-\rho^2)f(s).$$

Since $t(s)$ is linear, it is enough to show that $-2\rho\sqrt{f(s)} + (1-\rho^2)f(s)$ is convex. Let
$$g(x) \triangleq -2\rho\sqrt{x} + (1-\rho^2)x,$$

for $x > 0$, we need to show $g(f(s))$ is convex. Since $f(s)$ is convex (Lemma 6), $g(x)$ is convex in the range $\{x : x > 0\}$ (this is easily obeerved by, $g''(x) = \frac{\rho}{2}x^{-3/2} > 0$), it is enough to show that $g(x)$ is non-decreasing in its domain, and by a well-known convex analysis result (see, e.g. [11][Page 85]) we will obtain that $h(s)$ is convex.

To show that $g(x)$ is non-decreasing in its domain, we see that $g'(x) \geq 0$ when $x \geq \left(\frac{\rho}{1-\rho}\right)^2$. Since $f(x) \geq \frac{4}{(1-\rho)^2} \geq \left(\frac{\rho}{1-\rho}\right)^2$ by Lemma 6, $g(x)$ is non-decreasing in the range of $f(x)$. ∎

### B. The Upper Bound

As noted in Section V, the proof of the upper bound follows by bounding the coverage region from outside by a conic whose boundary is given by the points,

$$\left(\frac{d}{2} + \sqrt{\lambda}d\cos(\theta)\sqrt{1 - a\sin^2(\theta)},\right.$$
$$\left.\sqrt{\gamma}d\sin(\theta)\frac{\sqrt{1 - a\sin^2(\theta)}}{\sqrt{1-a}}\right) \quad (45)$$

We let $\mathcal{B}$ denote this conic. The proof follows along lines similar to those of Appendix D-A. Once again, we begin by focusing on the boundary of the conic $\mathcal{B}$, and show that on every point on this, $C_{\mathrm{DF}} \leq R$. We will later show that this implies that exterior of the conic is not contained in the coverage region either.

We must therefore show that at each point $\mathbf{a}_3$ on the boundary of $\mathcal{B}$,

$$C_{\mathrm{DF}}(P_2 = P_1, \alpha = 2) \leq R \quad (46)$$

where $C_{\mathrm{DF}}$ is given by (12).

Note that at a point (45) the boundary of $\mathcal{B}$, the following holds,

$$d_{13}^2 = d^2\left[\left(\frac{1}{2} + \sqrt{\lambda}\cos(\theta)\sqrt{1 - a\sin^2(\theta)}\right)^2\right.$$
$$\left. + \left(\sqrt{\gamma}\sin(\theta)\frac{\sqrt{1 - a\sin^2(\theta)}}{\sqrt{1-a}}\right)^2\right] \quad (47)$$

$$d_{23}^2 = d^2\left[\left(\frac{1}{2} - \sqrt{\lambda}\cos(\theta)\sqrt{1 - a\sin^2(\theta)}\right)^2\right.$$
$$\left. + \left(\sqrt{\gamma}\sin(\theta)\frac{\sqrt{1 - a\sin^2(\theta)}}{\sqrt{1-a}}\right)^2\right] \quad (48)$$

As in Appendix D-A, we may assume without loss of generality that $\rho$ in (12) is given by (3). With this choice, (46) becomes (recalling $P_1 = P_2$),

$$\log\left(1 + \frac{P_1}{d_{13}^2} + \frac{P_1}{d_{23}^2} + \frac{2\rho P_1}{d_{13}d_{23}}\right) \leq R = \log\left(1 + \frac{P_1}{d_{12}^2}(1-\rho^2)\right).$$

Substituting (47) and (48) into this equation, we get, after some manipulations, (46) becomes (49) at the top of the page. We now define $s \triangleq \sin^2(\theta)$, $t(s) \triangleq \frac{1}{4} + \lambda(1-as)$ and $f(s) \triangleq t^2(s) - \lambda(1-s)(1-as)$. With this notation, (49) becomes,

$$\frac{2t(s)}{f(s)} + \frac{2\rho}{\sqrt{f(s)}} - (1-\rho^2) \leq 0.$$

We will soon show (Lemma 9) that $f(s) > 0$, and thus, multiplying above equation by $f(s)/2$, we obtain that it is enough to show that

$$h(s) \triangleq t(s) + \rho\sqrt{f(s)} - \frac{1-\rho^2}{2}f(s) \leq 0$$

for all $s \in [0, 1]$. We will soon show (Lemma 10) that $h(s) = 0$ on the end points of the interval $s \in [0, 1]$; hence it is enough to prove that $h(s)$ is convex which will follow from Lemma 10.

The above discussion implies that the boundary of $\mathcal{B}$, $C_{\mathrm{DF}} \leq R$, with $C_{\mathrm{DF}} = R$ only when $\sin^2\theta = 0$ or 1. We now argue that this implies that the exterior of $\mathcal{B}$ lies outside $\mathcal{G}_{\mathrm{DF}}$. The discussion follows closely in the lines of a similar discussion in Appendix D-A. Consider an arbitrary point $\mathbf{a}_3$, and consider its polar coordinates (15). We also let $x_{\mathcal{B}}(\theta)$ denote the value of $x$ that coincides with the boundary of $\mathcal{B}$ (45). For all, $x \leq d$, we have, as in Appendix D-A, $C_{\mathrm{DF}} \geq R$. For $\theta$ such that $s = \sin^2\theta \notin \{0, 1\}$, we have shown above that $C_{\mathrm{DF}} < R$, and so $x_{\mathcal{B}}(\theta) > d$. At the other values of $\theta$, this can be shown by direct analytic calculation using (45). In the range $x > d$, $C_{\mathrm{DF}}$ can be shown to be strictly descending (as in Appendix D-A). Using our results for the boundary of $\mathcal{B}$, namely that $C_{\mathrm{DF}} \leq R$ when $x = x_{\mathcal{B}}(\theta)$, we obtain that for all $x > x_{\mathcal{B}}(\theta)$, $C_{\mathrm{DF}} < R$, and thus all such points (which constitute the exterior of $\mathcal{B}$) lie outside of $\mathcal{G}_{\mathrm{DF}}$.

We now prove some results that were required in the above discussion.

*Lemma 9:*  1) $\lambda - \gamma \leq 1$
2) $f''(s) \leq 0$
3) $f'(s) \leq 0$
4) $f(s) > 0$

*Proof:*
1)

$$\lambda - \gamma - 1$$
$$= \frac{1}{4} + \frac{1}{1-\rho} + \frac{\sqrt{2(1+\rho)}}{1-\rho^2} - \frac{2}{1-\rho} + \frac{1}{4} - 1$$
$$= \frac{2\sqrt{2(1+\rho)} - 2(1+\rho) - 1 + \rho^2}{2(1-\rho^2)}$$
$$= \frac{2\sqrt{2(1+\rho)} - 3 - 2\rho + \rho^2}{2(1-\rho^2)}$$

It is easy to see that $2\sqrt{2(1+\rho)} - 3 - 2\rho + \rho^2 \leq 0$ by proving convexity and checking the boundary points which completes the proof.

2) $f''(s) = 2(a\lambda - 1)a\lambda = 2(\lambda - \gamma - 1)a\lambda \leq 0$ by first part of this Lemma.

3) $f'(s) = 2t(s)t'(s) + \lambda(1-as) + a\lambda(1-s)$. Since $f''(s) \leq 0$ by the previous part of this Lemma, it is enough to prove $f'(0) \leq 0$. Or, $f'(0) = \lambda\left(\frac{3}{2} - \frac{\gamma}{2\lambda} - 2(\lambda - \gamma)\right) \leq$

$$\frac{1}{\frac{1}{4}+\lambda(1-a\sin^2(\theta))+\sqrt{\lambda(1-a\sin^2(\theta))}\cos(\theta)}+\frac{1}{\frac{1}{4}+\lambda(1-a\sin^2(\theta))-\sqrt{\lambda(1-a\sin^2(\theta))}\cos(\theta)}+ \quad (49)$$

$$\frac{2\rho}{\sqrt{\frac{1}{4}+\lambda(1-a\sin^2(\theta))+\sqrt{\lambda(1-a\sin^2(\theta))}\cos(\theta)}\sqrt{\frac{1}{4}+\lambda(1-a\sin^2(\theta))-\sqrt{\lambda(1-a\sin^2(\theta))}\cos(\theta)}} \leq 1-\rho^2$$

0. Hence, it is enough to prove that $3\lambda - \gamma \leq 4\lambda(\lambda - \gamma)$. After substituting the values of $\lambda$ and $\gamma$ and some manipulations, it reduces to proving that $\frac{2}{1+\rho} - \frac{(1-\rho)^2}{8} \geq 1$ which is true since left hand side decreases with $\rho \in [0,1)$ from $15/8$ at $\rho = 0$ to $1$ as $\rho \to 1$.

4) As $f'(s) \leq 0$ by the previous part of this Lemma, it is enough to prove $f(1) > 0$.

$$f(1) = t^2(1) = \left(\frac{1}{4}+\lambda(1-a)\right)^2 = \left(\frac{1}{4}+\gamma\right)^2 > 0.$$

∎

*Lemma 10:*
1) $h(0) = 0$
2) $h(1) = 0$
3) $h'''(s) \leq 0$
4) $h''(s) \geq 0$

*Proof:*

1) Proving this is equivalent to prove $\frac{2t(0)}{f(0)} + \frac{2\rho}{\sqrt{f(0)}} - (1-\rho^2) = 0$

$$\frac{2t(0)}{f(0)} + \frac{2\rho}{\sqrt{f(0)}} - (1-\rho^2)$$
$$= \frac{2(\lambda+1/4)}{(\lambda-1/4)^2} + \frac{2\rho}{\lambda-1/4} - (1-\rho^2)$$
$$= \frac{2(\lambda-1/4)+1}{(\lambda-1/4)^2} - (1-\rho^2)(1-\frac{2\rho}{1+\rho+\sqrt{2(1+\rho)}})$$
$$= \frac{2}{\lambda-1/4} + \frac{1}{(\lambda-1/4)^2} - \frac{1}{\lambda-1/4}(1-\rho+\sqrt{2(1+\rho)})$$
$$= \frac{1}{(\lambda-1/4)^2}\left((1+\rho-\sqrt{2(1+\rho)})(\lambda-1/4)+1\right)$$
$$= \frac{1}{(\lambda-1/4)^2}((1+\rho-\sqrt{2(1+\rho)})\frac{(1+\rho+\sqrt{2(1+\rho)})}{1-\rho^2}+1)$$
$$= \frac{1}{(\lambda-1/4)^2}(-1+1)$$
$$= 0$$

2) Proving this is equivalent to proving, $\frac{2t(1)}{f(1)} + \frac{2\rho}{\sqrt{f(1)}} - (1-\rho^2) = 0$

$$\frac{2t(1)}{f(1)} + \frac{2\rho}{\sqrt{f(1)}} - (1-\rho^2)$$
$$= \frac{2t(1)}{t^2(1)} + \frac{2\rho}{t(1)} - (1-\rho^2)$$
$$= 2\frac{1+\rho}{t(1)} - (1-\rho^2)$$
$$= 2\frac{1+\rho}{1/4+\gamma} - (1-\rho^2)$$
$$= 2\frac{1+\rho}{\frac{2}{1-\rho}} - (1-\rho^2)$$
$$= 0$$

3)
$$h(s) = t(s) + \rho\sqrt{f(s)} - \frac{1-\rho^2}{2}f(s)$$

$$h'(s) = t'(s) + \frac{\rho f'(s)}{2\sqrt{f(s)}} - \frac{1-\rho^2}{2}f'(s)$$

$$h''(s) = t''(s) + \frac{\rho}{2(f(s))^{(3/2)}}\left(f(s)f''(s) - \frac{(f'(s))^2}{2}\right) - \frac{1-\rho^2}{2}f''(s)$$

As $t(s)$ is linear,

$$h''(s) = \frac{\rho}{2(f(s))^{(3/2)}}\left(f(s)f''(s) - \frac{(f'(s))^2}{2}\right) - \frac{1-\rho^2}{2}f''(s)$$

$$h'''(s) = \frac{-3\rho}{4f(s)^{5/2}}f'(s)(f(s)f''(s) - f'(s)^2/2)$$

By Lemma 9, we see that $h'''(s) \leq 0$

4) Using $h'''(s) \leq 0$, it is enough to show that $h''(1) \geq 0$.
$f'(1) = -\frac{1}{4} - \frac{4}{(1-\rho)^2}\left(\sqrt{\frac{2}{1+\rho}}-1\right)$ $f''(1) = 2\left(-\frac{1}{4} + \frac{1}{(1-\rho)^2} + \frac{2}{(1-\rho^2)(1-\rho)} - 2\frac{\sqrt{2(1+\rho)}}{(1-\rho^2)(1-\rho)}\right)$
$h''(1) = \frac{1-\rho}{4}\left(-f''(1)(2+\rho) - \frac{\rho}{8}(1-\rho)^2 f'(s)^2\right)$
After some manipulations, proving $h''(1) \geq 0$ is equivalent to proving $(1-\rho)^2 - 4(1+\rho) - 8 + 8\sqrt{2(1+\rho)} + \frac{3}{4}\rho(1-\rho)^2 - \frac{\rho}{128}(1-\rho)^4 - \frac{\rho}{4}(1-\rho)^2\sqrt{\frac{2}{1+\rho}} \geq 0$
Hence, we will be done if we prove

a) $(1-\rho)^2 - 4(1+\rho) - 8 + 8\sqrt{2(1+\rho)} \geq 0$

b) $\frac{3}{4}\rho(1-\rho)^2 - \frac{\rho}{128}(1-\rho)^4 - \frac{\rho}{4}(1-\rho)^2\sqrt{\frac{2}{1+\rho}} \geq 0$

which are true since 1) is monotonically increasing with $\rho$ and 2) is monotonically decreasing with $\rho$, and the values at boundary points are satisfied.

∎

## APPENDIX E
## PROOF OF THEOREM 4

The proof follows closely in the lines of Appendix D-A. We review its main points below. As in that appendix, this proof follows by bounding the coverage region from within by an ellipse whose boundary is given by the points (38), but with $\lambda$ and $\gamma$ as defined by (9) and (10). We once again let $\mathcal{A}$ denote this ellipse, and begin by focusing on the boundary of $\mathcal{A}$, and show that it lies within the DF coverage region. By a similar discussion to that provided in Appendix D-A, this will imply that the interior of $\mathcal{A}$ is contained in the coverage region as well.

We must therefore show that at each point $\mathbf{a}_3$ on the boundary of $\mathcal{A}$,

$$C_{\text{DF}}(P_2 = P_1, \alpha = 4) \geq R \quad (50)$$

As in Appendix D-A, we begin by observing Note that by (38), at a point on the boundary of $\mathcal{A}$, $d_{13}$ and $d_{13}$ satisfy (40) and (41).

Once again, as in Appendices D-A and D-B, we may assume without loss of generality that $\rho$ in (12) is given by (8). With this choice, (50) becomes (recalling $P_1 = P_2$),

$$\log\left(1 + \frac{P_1}{d_{13}^4} + \frac{P_1}{d_{23}^4} + \frac{2\rho P_1}{d_{13}^2 d_{23}^2}\right) \geq R = \log\left(1 + \frac{P_1}{d_{12}^4}(1-\rho^2)\right).$$

Substituting (40) and (41) into this equation, after some manipulations, we obtain (51) at the top of the page.

We define $s, t(s)$ and $f(s)$ as in Appendix D-A (see (43) and (44)). Using this notation, (51) becomes,

$$\frac{2(f(s) + 2\lambda(1-s))}{f^2(s)} + \frac{2\rho}{f(s)} - (1-\rho^2) \geq 0.$$

We will soon show (Lemma 12) that $f(s) > 0$, and thus, multiplying above equation by $-\frac{f^2(s)}{2(1+\rho)}$, we obtain,

$$h(s) \triangleq (1-\rho)\frac{f^2(s)}{2} - f(s) - \frac{2\lambda(1-s)}{1+\rho} \leq 0$$

for all $s \in [0, 1]$. We will soon show (Lemma 11) that $h(s)$ is zero on the end points of the interval $s \in [0, 1]$; hence it is enough to prove that $h(s)$ is convex. This will soon be proved in Lemma 13.

The above discussion implies that the boundary of $\mathcal{A}$ lies within $\mathcal{G}_{\text{DF}}$. As in Appendix D-A, this implies that the interior lies within $\mathcal{G}_{\text{DF}}$ as well.

We now prove some results that were required in the above discussion.

*Lemma 11:*
1) $h(0) = 0$
2) $h(1) = 0$

*Proof:*
1)
$$\begin{aligned}
h(0) &= (1-\rho)\frac{f^2(0)}{2} - f(0) - \frac{2\lambda}{1+\rho} \\
&= (1-\rho)\frac{(t^2(0)-\lambda)^2}{2} - (t^2(0) - \lambda) - \frac{2\lambda}{1+\rho} \\
&= (1-\rho)\frac{(\lambda-\frac{1}{4})^4}{2} - (\lambda - \frac{1}{4})^2 - \frac{2(\lambda-\frac{1}{4})}{1+\rho} - \frac{1}{2(1+\rho)} \\
&= \frac{1-\rho}{2}[(\lambda-\frac{1}{4})^4 - \frac{2}{1-\rho}(\lambda-\frac{1}{4})^2 - \\
&\quad \frac{4}{1-\rho^2}(\lambda-\frac{1}{4}) - \frac{1}{(1-\rho^2)}] \\
&= \frac{1-\rho}{2}[0] \\
&= 0
\end{aligned}$$

2)
$$\begin{aligned}
h(1) &= (1-\rho)\frac{f^2(1)}{2} - f(1) \\
&= (1-\rho)\frac{t^4(1)}{2} - t^2(1) \\
&= t^2(1)(1 - \frac{1-\rho}{2}t^2(1)) \\
&= t^2(1)[1 - \frac{1-\rho}{2}(\gamma + \frac{1}{4})^2] \\
&= t^2(1)[1 - \frac{1-\rho}{2}\frac{2}{1-\rho}] \\
&= t^2(1)[1 - 1] \\
&= 0
\end{aligned}$$

∎

*Lemma 12:*
1) $f(s)$ is convex
2) $f'(s) \leq 0$ for all $s \in [0, 1]$
3) $f(s) \geq \frac{2}{1-\rho} > 0$

*Proof:*
1) $f''(s) = 2t'^2(s) = 2(\lambda - \gamma)^2 \geq 0$.
2) As $f''(s) \geq 0$, it is enough to prove $f'(1) \leq 0$.

$$\begin{aligned}
f'(1) &= 2t(1)t'(1) - \lambda(-1) \\
&= -2(\frac{1}{4} + \gamma)(\lambda - \gamma) + \lambda \\
&= -2\sqrt{\frac{2}{1-\rho}}(\lambda - \sqrt{\frac{2}{1-\rho}} + \frac{1}{4}) + \lambda
\end{aligned}$$

To prove $f'(1) \leq 0$, we need to prove $-2\sqrt{\frac{2}{1-\rho}}(\lambda - \sqrt{\frac{2}{1-\rho}} + \frac{1}{4}) + \lambda \leq 0$, or

$$\lambda - \frac{1}{4} \geq \frac{1}{\sqrt{1-\rho}}[\frac{1}{4}(2\sqrt{2} - \sqrt{1-\rho}) + \frac{2}{2\sqrt{2} - \sqrt{1-\rho}}]$$

Let $d = \frac{1}{4}(2\sqrt{2} - \sqrt{1-\rho}) + \frac{2}{2\sqrt{2}-\sqrt{1-\rho}}$. It is enough to show that

$$(\frac{d}{\sqrt{1-\rho}})^4 - \frac{2}{1-\rho}(\frac{d}{\sqrt{1-\rho}})^2 - \frac{4}{1-\rho^2}\frac{d}{\sqrt{1-\rho}} - \frac{1}{1-\rho^2} \leq 0$$

$$\frac{1}{\left(\frac{1}{4}+\lambda\cos^2\theta+\gamma\sin^2\theta-\sqrt{\lambda}\cos\theta\right)^2}+\frac{1}{\left(\frac{1}{4}+\lambda\cos^2\theta+\gamma\sin^2\theta+\sqrt{\lambda}\cos\theta\right)^2}$$
$$+\frac{2\rho}{\left(\frac{1}{4}+\lambda\cos^2\theta+\gamma\sin^2\theta-\sqrt{\lambda}\cos\theta\right)\left(\frac{1}{4}+\lambda\cos^2\theta+\gamma\sin^2\theta+\sqrt{\lambda}\cos\theta\right)}\geq 1-\rho^2 \quad (51)$$

Let $q = 2\sqrt{2} - \sqrt{1-\rho}$. It is enough to prove $d^2(d^2-2)(1+\rho) - 4d\sqrt{1-\rho} - (1-\rho) \le 0$. It is enough to prove $d(d^2-2)(1+\rho) - 4\sqrt{1-\rho} \le 0$, or $(\frac{4}{q^2} - \frac{q^2}{16})(\frac{2}{q} - \frac{q}{4})(1+\rho) \le 4\sqrt{1-\rho}$. Hence, we will be done if we prove

a) $\frac{4}{q^2} - \frac{q^2}{16} \le 1$
b) $\frac{2}{q} - \frac{q}{4} \le 2\sqrt{1-\rho}$
c) $1 + \rho \le 2$

where (a) follows since $q^2 \ge (2\sqrt{2}-1)^2 \ge 8(\sqrt{2}-1)$, Hence, $q^4 + 16q^2 - 64 \ge 0$, (b) is equivalent to $\sqrt{1-\rho} \le \frac{12}{7}\sqrt{2}$ which is true and (c) follows since $\rho \le 1$.

3) As $f'(s) \le 0$ (by second part of this Lemma), $f(s) \ge f(1) = t^2(1) = \frac{2}{1-\rho}$.

∎

*Lemma 13:* $h(s)$ is convex.

*Proof:*
$$h(s) \triangleq (1-\rho)\frac{f^2(s)}{2} - f(s) - \frac{2\lambda(1-s)}{1+\rho}$$

As $\frac{2\lambda(1-s)}{1+\rho}$ is linear, it is enough to show that $(1-\rho)\frac{f^2(s)}{2} - f(s)$ is convex. Let
$$g(x) \triangleq (1-\rho)\frac{x^2}{2} - x,$$

we need to show $g(f(s))$ is convex. Since $g(x)$ is convex ($g''(x) = 1-\rho > 0$) and $f(s)$ is convex (Lemma 12), it is enough to show that $g(x)$ is non-decreasing in its domain, and as in the proof of Lemma 8, we will obtain that $h(s)$ is convex.

To show that $g(x)$ is non-decreasing in in the range of $f(x)$, we see that $g'(x) \ge 0$ when $x \ge \frac{1}{1-\rho}$. Since $f(x) \ge \frac{2}{1-\rho} > \frac{1}{1-\rho}$ by Lemma 12, $g(x)$ is non-decreasing in the range of $f(x)$.

∎

## APPENDIX F
## PROOF OF LEMMA 1

$$\pi\sqrt{\lambda\gamma}d^2 = \pi\sqrt{\lambda\gamma(1-\rho^2)}d_c^2$$
$$= \pi d_c^2\sqrt{(\lambda\sqrt{1-\rho})(\gamma\sqrt{1-\rho})(1+\rho)} \quad (52)$$

We now variables from $d$ to $\rho$, using (8). Taking $d \to 0$ is equivalent to taking $\rho \to 1$.

$$\lim_{\rho\to 1}\gamma\sqrt{1-\rho} = \sqrt{2} \quad (53)$$

By definition of $\lambda$ (Theorem 4), $\lambda\sqrt{1-\rho}$ is the largest solution of
$$\left(\frac{t}{\sqrt{1-\rho}} - \frac{1}{4}\right)^4 - \frac{2}{1-\rho}\left(\frac{t}{\sqrt{1-\rho}} - \frac{1}{4}\right)^2$$
$$-\frac{4}{1-\rho^2}\left(\frac{t}{\sqrt{1-\rho}} - \frac{1}{4}\right) - \frac{1}{1-\rho^2} = 0$$

Multiplying by $(1-\rho)^2$
$$\left(t - \frac{\sqrt{1-\rho}}{4}\right)^4 - 2\left(t - \frac{\sqrt{1-\rho}}{4}\right)^2 -$$
$$\frac{4\sqrt{1-\rho}}{1+\rho}\left(t - \frac{\sqrt{1-\rho}}{4}\right) - \frac{1-\rho}{1+\rho} = 0$$

Thus, $\lambda\sqrt{1-\rho} = \sqrt{1-\rho}/4 + \psi$, where $\psi$ is the largest solution of
$$t^4 - 2t^2 - \frac{4\sqrt{1-\rho}}{1+\rho}t - \frac{1-\rho}{1+\rho} = 0$$

It is easy to see that as $\rho \to 1$, the largest solution of above equation approaches $\sqrt{2}$ (roots vary continuously with the coefficients of the polynomial, hence the limit of the root of the polynomial equals the root of the limiting polynomial [12]). Hence,
$$\lim_{\rho\to 1}\lambda\sqrt{1-\rho} = \lim_{\rho\to 1}\frac{\sqrt{1-\rho}}{4} + \sqrt{2} = \sqrt{2} \quad (54)$$

Substituting (53) and (54) in (52), we obtain,
$$\lim_{d\to 0}\pi\sqrt{\lambda\gamma}d^2 = 2\pi d_c^2.$$

We now consider the true coverage region when $d \to 0$. Taking $d \to 0$ is equivalent to placing the relay at the source. Examining (12), it is straightforward to observe that this produces a coverage region which is a circle with radius $\sqrt{2}d_c$. Thus, the true area coincides with the above computed limit of the lower bound, when $d \to 0$.

∎

## APPENDIX G
## PROOF OF THEOREM 5

The proof of this theorem follows along lines similar to those of Theorem 1. The half-duplex DF and CF achievable rates were computed by Zhixin *et al.* [10], and are given by,

$$C_{\text{DF}} = \max_{0\le\rho\le 1}\min\left\{t\log\left(1 + \frac{P_1}{d_{12}^\alpha}\right) + (1-t)\times\right.$$
$$\log\left(1 + (1-\rho^2)\frac{P_1}{d_{13}^\alpha}\right), t\log\left(1 + \frac{P_1}{d_{13}^\alpha}\right) +$$
$$\left.(1-t)\log\left(1 + \frac{P_1}{d_{13}^\alpha} + \frac{P_2}{d_{23}^\alpha} + \frac{2\rho\sqrt{P_1P_2}}{(d_{13}d_{23})^{\alpha/2}}\right)\right\} \quad (55)$$

$$C_{\text{CF}} = t\log\left(1+\frac{P_1}{d_{13}^\alpha}+\frac{P_1}{d_{12}^\alpha(1+\hat{N}_2)}\right)$$
$$+(1-t)\log\left(1+\frac{P_1}{d_{13}^\alpha}\right) \quad (56)$$

where

$$\hat{N}_2 = \frac{1+P_1(\frac{1}{d_{12}^\alpha}+\frac{1}{d_{13}^\alpha})}{(1+\frac{P_1}{d_{13}^\alpha})\left((1+\frac{P_2/d_{23}^\alpha}{1+P_1/d_{13}^\alpha})^{(1-t)/t}-1\right)} \quad (57)$$

We begin by proving $\mathcal{G}_{\text{CF}}(d) \supseteq \mathcal{G}_{\text{NR}}(d)$, for all $d > 0$. This is straightforward from the observation that by (56) and (16), $C_{\text{CF}} > C_{\text{NR}}$ at every destination point $\mathbf{a}_3$.

We now prove Part 2 of the theorem, which (as in the proof of Theorem 1) is the easier of the two parts. To show that $\mathcal{G}_{\text{NR}} > \mathcal{G}_{\text{DF}}(d)$, we first note that by (16) and (1), it can easily be observed that $\mathcal{G}_{\text{NR}}$ is exactly a sphere with radius $d_c$. We now show that DF cannot support a rate of $R$ outside this sphere.

Let $\theta \in [0, 2\pi)$, $x > d_c$, and consider the achievable rate $C_{\text{DF}}$ at the point $\mathbf{a}_3$ whose polar coordinates (see Appendix A-B) are $(\theta, x)$,

$$C_{\text{DF}} \stackrel{(a)}{\leq} t\log\left(1+\frac{P_1}{d_{12}^\alpha}\right)+(1-t)\log\left(1+\frac{P_1}{d_{13}^\alpha}\right)$$
$$\stackrel{(b)}{<} t\log\left(1+\frac{P_1}{d_c^\alpha}\right)+(1-t)\log\left(1+\frac{P_1}{d_c^\alpha}\right)$$
$$= \log\left(1+\frac{P_1}{d_c^\alpha}\right)$$
$$\stackrel{(c)}{=} R \quad (58)$$

where (a) follows from (55), (b) follows since $x > d_c$, $d > d_c$, and (c) follows from (1). Thus, $C_{\text{DF}}$ is strictly less that $R$ outside the sphere of radius $d_c$, as desired.

To prove the first part of the theorem, we again apply the notation $x_S(\theta)$ that was defined in Appendix B. By (55), we may obtain a lower bound on $C_{\text{DF}}$ by restricting the maximization to $\rho = 0$. Thus (recall that we have assumed $t = 1/2$),

$$C_{\text{DF}} \geq \frac{1}{2}\min\left\{\log\left(1+\frac{P_1}{d_{12}^\alpha}\right)+\log\left(1+\frac{P_1}{d_{13}^\alpha}\right),\right.$$
$$\left.\log\left(1+\frac{P_1}{d_{13}^\alpha}\right)+\log\left(1+\frac{P_1}{d_{13}^\alpha}+\frac{P_2}{d_{23}^\alpha}\right)\right\}$$
$$= \frac{1}{2}\log\left(1+\frac{P_1}{d_{13}^\alpha}\right)+\frac{1}{2}\min\left\{\log\left(1+\frac{P_1}{d_{12}^\alpha}\right),\right.$$
$$\left.\log\left(1+\frac{P_1}{d_{13}^\alpha}+\frac{P_2}{d_{23}^\alpha}\right)\right\} \quad (59)$$

The following lemma examines this expression at $x = x_{\text{DF}}(\theta)$.

*Lemma 14:* For all $\theta$, consider the point $\mathbf{a}_3$ whose polar coordinates are given by $(\theta, x_{\text{DF}}(\theta))$. Then (18) holds at $\mathbf{a}_3$,

*Proof:* Assume, by contradiction, that (18) does not hold. Letting $x = x_{\text{DF}}(\theta)$, we will now show (as in the proof of Lemma 3) that we may increase $x$ and preserve $C_{\text{DF}} \geq R$, contradicting the maximality of $x_{\text{DF}}(\theta)$.

Consider (55). Whenever (18) does not hold, we have by (59),

$$C_{\text{DF}} \geq \frac{1}{2}\log\left(1+\frac{P_1}{d_{13}^\alpha}\right)+\frac{1}{2}\log\left(1+\frac{P_1}{d_{12}^\alpha}\right) > \frac{1}{2}\log\left(1+\frac{P_1}{d_{12}^\alpha}\right)$$
$$\stackrel{(a)}{\geq} \frac{1}{2}\log\left(1+\frac{P_1}{d_c'^\alpha}\right) \stackrel{(b)}{=} R$$

Where the (a) was obtained by the fact that $d_{12} = d$ and $d \leq d_c'$, and (b) was obtained by (11). Since $C_{\text{DF}} > R$, by a continuity argument, we may increase $x$ slightly and still get $C_{\text{DF}} \geq R$. Therefore, if $C_{\text{DF}} > R$ at $x = x_{\text{DF}}(\theta)$, it will remain so after we increase $x$ a little, producing the desired contradiction. ∎

By Lemma 14, (17) implies that at $x = x_{\text{DF}}(\theta)$,

$$C_{\text{DF}} \geq \frac{1}{2}\log\left(1+\frac{P_1}{d_{13}^\alpha}\right)+\frac{1}{2}\log\left(1+\frac{P_1}{d_{13}^\alpha}+\frac{P_2}{d_{23}^\alpha}\right)$$
$$> C_{\text{CF}}$$

The last inequality is obtained by simple arithmetic, using (56) (recalling that $t = 1/2$) in a similar way to the way (20) was obtained in Appendix B. Again, by similar arguments as in Appendix B, this implies that $C_{\text{CF}} < R$ for all $x > x_{\text{DF}}(\theta)$, and thus $\mathcal{G}_{\text{DF}}(d) \supset \mathcal{G}_{\text{CF}}(d)$ as desired. This completes the proof of Part 1 of the theorem. ∎

## APPENDIX H
## EXTENSIONS TO PHASE FADING

### A. Extension of Theorem 1 to Phase Fading

The achievable rates with DF and CF and the upper bound, which were computed in [5] are given by,

$$C_{\text{DF}} = \min\left\{\log\left(1+\frac{P_1}{d_{12}^\alpha}\right),\right.$$
$$\left.\log\left(1+\frac{P_1}{d_{13}^\alpha}+\frac{P_2}{d_{23}^\alpha}\right)\right\} \quad (60)$$

$C_{\text{CF}}$ is as given in Appendix A, expressions (13) and (14).

$$C_{\text{UB}} = \min\left\{\log\left(1+P_1(\frac{1}{d_{12}^\alpha}+\frac{1}{d_{13}^\alpha})\right),\right.$$
$$\left.\log\left(1+\frac{P_1}{d_{13}^\alpha}+\frac{P_2}{d_{23}^\alpha}\right)\right\} \quad (61)$$

The proof follows closely in the lines of Appendix B. The expressions for $C_{\text{CF}}$ and $C_{\text{NR}}$ remain unchanged in the phase fading setting and thus the result $\mathcal{G}_{\text{CF}}(d) \supseteq \mathcal{G}_{\text{NR}}(d)$ carries over immediately. Examining (60), it is easily observed to be equal to (12) with the maximization replaced with an assignment of $\rho = 0$. The remainder of the proof follows along direct lines as Appendix B and is omitted.

## B. Proof of Lemma 2

Lemma 3 (Appendix B) can straightforwardly be shown to carry over to the phase-fading setting. With this lemma, at the point whose polar coordinates are $(\theta, x_{\text{DF}}(\theta))$ (for arbitrary $\theta$), (61) reduces to $C_{\text{UB}} = \log\left(1 + P_1/d_{13}^\alpha + P_2/d_{23}^\alpha\right)$, which coincides with $C_{\text{DF}}$ at that point. By continuity arguments, $C_{\text{DF}}$ must equal $R$ at $(\theta, x_{\text{DF}}(\theta))$ and thus $C_{\text{UB}}$ equals $R$ as well.

In Appendix B we showed that the value of $C_{\text{CF}}$ at a point whose polar coordinates are $(\theta, x)$, descends as a function of $x$ for $x > x_{\text{DF}}(\theta)$. Similar arguments can be now be applied to $C_{\text{UB}}$, again rendering $\mathcal{G}_{\text{UB}}(d) \subseteq \mathcal{G}_{\text{DF}}(d)$. Since we trivially have $\mathcal{G}_{\text{UB}}(d) \supseteq \mathcal{G}_{\text{DF}}(d)$, this completes the proof of the lemma. ∎